# Synthesis and Characterizations of $CH_3NH_3PbI_3$:ZnS Microrods for Optoelectronic Applications

Mohammad Tanvir Ahmed[1], Shariful Islam[1*], Muhammad Shahriar Bashar[2], Md. Abul Hossain[1], Farid Ahmed[1]

[1] Department of Physics, Jahangirnagar University, Dhaka-1342, Bangladesh.

[2] IFRD, Bangladesh Council of Scientific and Industrial Research, Dhanmondi, Dhaka-1205, Bangladesh.

*Corresponding Author. Email: s_islam@juniv.edu

**Graphical Abstract**

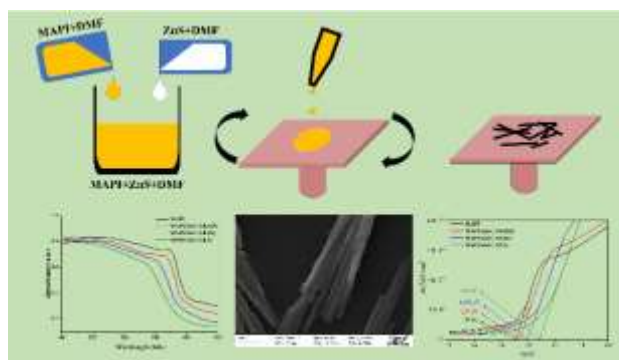

**Abstract**

Organometallic perovskite is one of the potential materials in the various optoelectronic research fields. This research work demonstrates the synthesis of $CH_3NH_3PbI_3$:ZnS microrods via one-step spin coating for optoelectronic applications. Incorporation of ZnS in the perovskite material caused bandgap variation in the visible wavelength range. Structural and chemical properties have been observed using X-ray diffraction, scanning electron microscope, and Fourier transform infrared spectroscopy. The optical and electrical properties were studied via Uv-vis spectroscopy and impedance analyzer. The addition of ZnS caused to increase the optical absorption coefficient from $\sim 10^4$ cm$^{-1}$ to $\sim 10^5$ cm$^{-1}$. The bandgaps of the thin films was calculated using the Tauc relation and electrical method, obtained in the range 1.51 eV – 1.64 eV, suitable for various optoelectronic applications.

## 1. Introduction

Organometallic perovskites (OMPs) are very multifunctional materials with widespread applications such as photodetectors, solar cells (SCs), field-effect transistors (FET), light-emitting diode, laser, light-emitting electrochemical cells, and so on [1]. They can offer a tunable bandgap and a better visible light absorption property, making them highly potential material for various optoelectronic (OE) applications. These plentiful applications led scientists all over the world to study the different properties of these materials. The material has also become very popular because of the easy synthesis process and low cost. The general formula of OMPs is $ABX_3$, where A, B, and X are monovalent organic cations (e.g., methylammonium, formamidinium, etc.), divalent metal cation (e.g., $Pb^{2+}$, $Sn^{2+}$, etc.), and halogen anion, respectively [2]. One of the greatest successes of OMPs in SCs technology is an efficient light absorber with a maximum conversion efficiency of 25.5%, which is reported by the national renewable energy laboratory [3].

The optical and electrical performance of a material highly depends on its crystal structure and bandgap. Depending on various crystal growth conditions, a material can offer different crystal structures (one, two, or three-dimensional). In the case of micro/nanorod structure, the bandgap also depends on the diameters. Hence changing the diameters can cause an alteration in their bandgap, which makes them potential materials for numerous applications. Perovskite microrods (MRs), nanorods (NRs), or nanowires (NWs) can provide better performances in different OE applications because of their unique crystal structure and tunable bandgap. One-dimensional (1D) OMPs have become quite popular due to better photovoltaic performance and other OE applications [4].

Im et al. have first synthesized the $CH_3NH_3PbI_3$ (MAPI) NWs in 2015 via two-step spin coating technique with a variation of DMF content [5]. They successfully demonstrated that the formation of MAPI NWs is more probable with a higher amount of DMF. The prepared NWs showed a fine absorption of visible wavelength, but the absorbance was slightly lower in comparison with bulk MAPI. In 2016, Spina et al. also synthesized NWs of MAPI via slip coating method for photo-detecting application [6]. This was a unique and one of the easiest methods for MAPI crystal growth. The process can avoid random nucleation hence control the crystal growth. They have not demonstrated the visible light absorption properties of this 1D MAPI. Wu et al. in 2018, have also studied the effect of DMF amount present in MAPI on the formation of 1D MAPI [7]. They successfully showed that at 1D MAPI could be obtained via increasing the DMF contents. They also demonstrated the higher absorption properties of OMP NWs than 3D MAPI. This research disproves the low absorption of 1D MAPI as reported by Im et al. [5]. In 2019, Mishra et al. synthesized MAPI MRs by cooling down MAPI precursor solution to room temperature [8]. They reported that MAPI MRs could join parallelly, forming a long wire. But the MRs growth process is rather slow, and no optical properties of the MRs had been observed for OE applications. Zhang et al. reported $CH_3NH_3PbBr_3$ MRs synthesis in 2019 through dip-coating technique for dynamically switchable micro-laser [9]. However, fine absorption was obtained only in the wavelength below 550nm.

Research on composite perovskites also got the attention of scientists. Various materials as composites have been added with pure perovskites structure to get better performance and stability. Among them, perovskite composite with metal sulfide is a new direction of research to enhance optical performance, which was started in 2017. Chen et al. reported a MAPI:CdS film for SCs prepared via precursor blending method which showed a better photovoltaic performance [10]. The absorbance was slightly increased between 500nm to 750nm in presence of CdS suggesting an increase in absorption coefficient. However, Cd is a highly toxic material and should be avoided. In 2019, Wang et al. synthesized PbS quantum dots embedded in MAPI NWs for photodetection application [11]. The composite nanorods showed better absorbance in a wide range of the visible spectrum than pure MAPI rods. These achievements uncovered that incorporating metal sulfide in OMPs can offer more enhanced optical performance, opening an alternative route in perovskite research and motivating us to conduct our research.

For deposition of thin film, numerous methods have been established, e.g., sol-gel spin coating, chemical vapor deposition (CVD), chemical bath, spray pyrolysis, and so on [12]. The sol-gel method can provide better crystallization with enhanced homogeneity at a relatively low temperature [13]. Though thermal evaporation, sputtering, CVD can provide better uniformity of thin-film, these methods are costly compared to spray pyrolysis, chemical bath, spin-coating, or doctor blade method. Among these economical methods, spin-coating can provide better homogeneity for small area of fabrication of thin film with minimal thickness [14]. Due to the simplicity and cost-effectivity, spin coating is a favourable deposition method of perovskites.

Here we report the synthesis of MAPI:ZnS MRs via one step spin coating method. ZnS is also a potential material for OE applications with a high absorption coefficient [15], which has been used as ETL in PSCs, showing better photovoltaic performance with a PCE of 17.4% [16]. To our best knowledge, various OE properties of MAPI:ZnS material have never been studied before. Here we demonstrated the growth of MAPI:ZnS MRs in thin films, and reported various structural, morphological, optical, etc. alterations due to the variation in ZnS stoichiometry.

## 2. Experimental Details

*2.1. Materials*

Methylamine (MA) solution 33wt% (Sigma-Aldrich), hydroiodic acid (HI) 57% (Sigma-Aldrich), $PbI_2$ 99% (Sigma-Aldrich), diethyl ether, N, N-Dimethylformamide (DMF), Zinc chloride (Merck), and thiourea (Merck) were used in this research process.

*2.2. Preparation of MAPI and MAPI:ZnS Crystals*

The synthesis process of $CH_3NH_3I$ (MAI) was followed by previous research [17]. MA and HI solution of (1:1) molar ratio was mixed and stirred in an ice bath for 1 hour and further kept in the bath for 2h. The solution was dried in an oven at 60ºC and the MAI crystal was formed, which was washed several times with diethyl ether to remove impurities. $PbI_2$ and MAI were dissolved completely in DMF at an equimolar ratio and dried at 60ºC to obtain MAPI crystals. **Figure-1** shows the prepared MAPI crystals.

The precursor blending solution method was used to prepare MAPI:ZnS crystals [10]. $ZnCl_2$ and thiourea were dissolved by a 1:1 molar ratio in the MAPI-DMF solution and stirred for 30 minutes and evaporated in the oven at 60ºC. MAPI:ZnS with three different molar ratios (e.g., 1:0.025, 1:0.05, 1:0.1) were prepared. The pH of the MAPI, MAPI:ZnS (1:0.025), MAPI:ZnS (1:0.025), and MAPI:ZnS (1:0.025) solutions are respectively 6.11, 6.11, 6.12, and 6.13.

*2.3. Thin-film preparation*

MAPI and MAPI:ZnS precursor solutions of 0.5gm/ml were prepared with DMF solvent. The viscosities of the MAPI, MAPI:ZnS (1:0.025), MAPI:ZnS (1:0.025), and MAPI:ZnS (1:0.025) solutions are measured to be 6.2822 mPa-s, 6.2782 mPa-s, 6.2731 mPa-s, and 6.2694 mPa-s, respectively. The Glass substrates were cleaned with deionized water, ethanol, and acetone via ultrasonic cleaner. Preheated substrates were spin-coated with a few drops of prepared solution at 1500 rpm for 30 sec and then annealed at 80°C for 30 minutes. **Figure-2** shows the prepared MAPI and MAPI:ZnS thin films.

*2.4. Characterization*

The X-ray Diffraction (XRD) and Fourier transform infrared (FTIR) spectroscopy analysis was performed for all the prepared samples via Explorer Diffractometer and IRPrestige-21, respectively, at Wazed Miah Science Research Centre (WMSRC), Jahangirnagar University (JU), Savar, Dhaka-1342, Bangladesh. The thin film thickness was measured by DektaXT-A Surface Profilometer at the Bangladesh Council of Scientific and Industrial Research (BCSIR), Dhanmondi, Dhaka-1205, Bangladesh. UV-vis spectroscopy and SEM were also performed at BCSIR via UH4150 Spectrophotometer and ZEISS EVO 18 SEM, respectively. Wayne Kerr 6500B Impedance Analyzer was used to measure the electrical resistance.

## 3. Result and Discussion

*3.1. XRD Analysis*

The XRD patterns for the samples MAPI and ZnS doped MAPI are shown in **Figure-3**. The diffraction peaks for MAPI were observed at 2θ values of 14.09º, 19.88º, 24.46º, 28.32º, 31.8º, 34.9º, 40.50º, and 43.07º, corresponds to (100), (110), (111), (200), (102), (112), (202), (212) planes respectively which satisfies previous report [18]. The addition of ZnS causes a structural deformation in the MAPI crystals which is observed from the spectra exhibiting a slight shifting of the diffraction peaks toward higher 2θ values and from the lattice parameters, shown in **Table-1**. The slight shifting toward higher 2θ indicates the increase of structural stress and decrease of interplanar spacing, which can cause the shrinkage of lattice [19]. No clear peaks of ZnS have been observed in the XRD spectra. However, an overlapped peak with the perovskite peak at 28.15º and a new peak at 32.68º are analogous with ZnS's corresponding peaks according to the Crystallography Open Database (COD ID- 1539414).

The absence of $ZnCl_2$ peaks can verify the formation of ZnS in the perovskite sample since ZnS was formed through the chemical reaction of $ZnCl_2$ and Thiourea ($ZnCl_2 + CH_4N_2S = ZnS + CH_4N_2 + Cl_2$). The variation of ZnS concentration in the sample causes deformation in the perovskite's unit cell structure, shown in **Table 1**. The overall cell volume of MAPI decreases with increasing ZnS concentration. The XRD peaks slightly broadened after ZnS incorporation, which represents a reduction in crystallite size and an increase of lattice defects [19].

*3.1.1. Crystallite Size and Lattice Strain*

A crystallite is a small region in a material made up of a single crystal, i.e., atomic regularity is exactly maintained in the region. On the other hand, lattice strain measures the distribution of lattice parameters resulting from imperfections, lattice dislocation in a crystal. Both crystallite size and lattice strain affect the Bragg peaks through broadening and shifting. The crystallite sizes and lattice strains are calculated from the **equation 1 -2** [20–22] and listed in **Table-2**.

$$D = \frac{\kappa \lambda}{\Gamma \cos \theta}, \qquad (1)$$

$$\xi = \frac{\Gamma}{4 \tan \theta}. \qquad (2)$$

Here $D$, $\xi$, $\Gamma$, $\kappa$, and $\lambda$ are crystallite size, lattice strain, full-width half maxima, Scherrer constant (=0.9), and X-ray wavelength [20–22]. From the calculations, it was observed that the average crystallite size decreased with the increase of ZnS concentration. The presence of Zn or S atoms in the perovskites can cause irregularities and defects during the MAPI crystal growth. This can cause the reduction of average crystallite size. The lattice strain increases with increasing ZnS concentration. The increased strain represents higher stress due to the incorporation of foreign atoms. This stress and strain can also rise due to lattice mismatch, i.e., dissimilarity in the lattice structure of MAPI and the substrate as well as due to the deferences in thermal expansion coefficients ($\alpha_L$) between the film and substrate ($\alpha_L$ for MAPI is 43.3-33.3 $\times 10^{-6}$ $K^{-1}$ and for the soda-lime substrate is $9 \times 10^{-6}$ $K^{-1}$) [23,24].

*3.1.2. Dislocation Density and Stacking Fault*

The dislocation density ($\rho_d$) represents the total dislocation length per unit volume in a crystalline material, which denotes irregularities and material strength. Generally, a material with a higher dislocation density offers increased strength [25]. On the other hand, stacking faults ($S_F$) are abnormalities in the crystal plane stacking sequence that violate the ideal lattice's regularity [26]. The dislocation density and stacking fault were calculated from equations 3-4 [27,28],

$$\rho_d = \frac{1}{D^2}, \qquad (3)$$

$$S_F = \frac{2\pi^2 \Gamma}{45\,(3\tan\theta)^{\frac{1}{2}}}. \qquad (4)$$

where D and $\Gamma$ are the crystallite size and full-width half maxima, respectively. Both $\rho_d$ and $S_F$ increase with ZnS concentration, which signifies that the presence of ZnS increases defects in the MAPI crystals growth. These increased defects can be identified as the cause of the reduction of average crystallite size. The increased value of $\rho_d$ represents lower crystallinity of the material, i.e., the overall crystallinity decreased after ZnS addition. This also signifies that MAPI:ZnS (1:0.1) possess higher strength compared to pure MAPI.

*3.2. FTIR Analysis*

FTIR spectra were observed in the wavenumber range 750 cm$^{-1}$ — 4000 cm$^{-1}$ (**Figure-5**) for pure perovskite and perovskite:ZnS crystals. The broad peak between 3455 cm$^{-1}$ — 3687 cm$^{-1}$ represents O-H stretching due to the presence of moisture in the samples. The peaks of very less intensity at 3187 cm$^{-1}$ is of $NH_3^+$ asymmetric stretching and at 3130 cm$^{-1}$ is of $NH_3^+$ symmetric stretching. $CH_3$ asymmetric and symmetric stretching were observed at 2958 cm$^{-1}$ and 2918 cm$^{-1}$, respectively [29]. The peaks at 1585 cm$^{-1}$, 1460 cm$^{-1}$, and 1422 cm$^{-1}$ are due to asymmetric $NH_3^+$ bending, symmetric $NH_3^+$ bending, and $CH_3$ bending, respectively [30]. The peak at 1020 cm$^{-1}$ represents C-N stretching [31]. The peaks at 1255 cm$^{-1}$ and 944 cm$^{-1}$ represent $CH_3$-$NH_3^+$ rocking [32]. Due to the transparency of $PbI_2$ to infrared wavelength, no peak for Pb-I was observed in the spectrum [33].

For small stoichiometry of the ZnS mixture in the sample, very slight changes in the peak position were observed. But after reaching the molar ratio of MAPI and ZnS to 1:0.1, all the peak position changes significantly due to the structural deformation. $NH_3^+$ stretching peaks were observed at 3193 cm$^{-1}$ (asymmetric) and 3139 cm$^{-1}$ (symmetric). The peaks at 2948 cm$^{-1}$, 1467 cm$^{-1}$, and 1010 cm$^{-1}$ define $CH_3$ stretching, $CH_3$ bending, and C-N stretching. Both peaks of $CH_3$-$NH_3^+$ rocking shifted to 1250 cm$^{-1}$ and 954 cm$^{-1}$. A new peak at 1099 cm$^{-1}$ arises with ZnS increment, which can be identified as the characteristic peak of ZnS [34]. Again, a strong peak at 1625 cm-1 was observed, previously characterized as the absorption of inorganic sulfide compounds [35].

*3.3. Surface Morphology*

**Figure-6** shows the SEM images of pure perovskite and perovskite:ZnS MRs in the thin films. MAPI crystals generally grow through a dendrite structure; this dendrite formation allows perovskite to extend in a particular direction uniformly, which finally takes the form of wires or rods [36]. MAPI MRs formation is observed in the thin film with an average rod diameter of about 612nm.

After the addition of ZnS, porosity was observed in the perovskite structures. The rod structure started to break down and transformed into the smaller irregular-sized grain with ZnS increment. Some of ZnS may have formed in the path of MAPI crystal growth which opposed the further crystal growth. Hence the grain size decreased and as well as porosity increased. In **Figure-6(c-d)** the change in grain structure with the increments of ZnS constituent in the samples are clearly observed. The average diameter of the irregular grains, finally observed, is 508nm.

*3.4. Elemental Analysis*

**Figure-7** shows the Energy-dispersive X-ray (EDX) spectra of the prepared thin films. The EDX analysis reveals the presence of C, N, Pb, I, Zn, and S atoms in the samples. In the targeted area of MAPI and MAPI:ZnS thin films, the atomic ratios of Pb/I are 0.42, 0.35, 0.32, and 0.30, which are close to previous reports [37,38]. According to a previous report, MAPI is very sensitive towards high energy electron beam and easily decomposes through the EDX analysis process, which causes variations in the Pb/I ratio [39].

The increment Zn and S atoms in the films is observed with the sequential order maintaining the atomic ratio of Zn/S between 0.83 and 0.97. The atomic % and weight % of the different elements observed in the EDX spectra are listed in **Table 3**.

*3.5. Optical Characteristics*

The UV-visible absorption was analyzed in the wavelength range of 400nm─900nm for each thin film. **Figure-8** shows the absorption spectrum of the prepared thin films. As observed in earlier reports, the MAPI film showed a fine absorption in the visible and near-infrared wavelength range [40,41]. The absorption onset for pure MAPI thin film is about 800 nm, consistent with the report of Tombe *et al.* [41]. The absorbance edge shifted to short-wavelength values as the ZnS concentration increased, suggesting that the bandgap energy of the produced perovskite thin films increased. The absorption spectrum follows blue shifting with further increasing ZnS concentration in the sample. The reason is that ZnS shows strong absorption in the short wavelength region [42,43].

The absorption coefficients ($\alpha$) shown in **Figure-9** of the thin films calculated using the following equation [44],

$$\alpha = \frac{2.303 A}{t}, \quad (5)$$

where A is the absorbance and t is the film thickness measured by surface profilometer. The MAPI film had an absorption coefficient above $10^4$ cm$^{-1}$, making it a potential candidate for numerous OE applications, especially for the SCs absorber layer [45]. The absorption coefficient increased with the increase in ZnS concentration and reached up to the order of $10^5$ cm$^{-1}$. The absorption coefficient edge is blue-shifted with the rise of ZnS concentration in MAPI thin film. This phenomenon occurred because ZnS has a high absorption coefficient in the low wavelength range [15,46]. The value of α for pure MAPI at 400 nm and 750 nm are $7.7 \times 10^4$ cm$^{-1}$ and $6.9 \times 10^4$ cm$^{-1}$, respectively. The electromagnetic wave of 400 nm and 750 nm have a penetration depth (δ) of 0.13 μm and 0.14 μm, respectively which means 37% of the incident 400 nm and 750 nm wavelength are absorbed after traversing 0.13 μm and 0.14 μm thickness of the film, respectively. On the other hand, for MAPI:ZnS (1:0.1) thin film, α is $1.08 \times 10^5$ cm$^{-1}$ and $4.1 \times 10^4$ cm$^{-1}$ for 400 nm and 750 nm wavelengths, respectively. The corresponding penetration depths are 0.1 μm and 0.25 μm, i.e., only a few microns thick films are sufficient to absorb most of the visible spectrum.

The reflectance spectra of the thin films are shown in **Figure-10**. The reflectance is very low (<7.5%) for all the samples and varies very slightly in the visible wavelength range. This less reflectance corresponds to minor energy loss due to reflection. The incorporation of ZnS caused to increase in the reflectance of the thin films. This observation represents the reduction of crystallite size and increase of imperfection since crystal imperfection can increase the scattering of photons [47]. Overall low reflectance indicates that the loss of incident energy due to reflection is very less, which is preferable to OE applications. These materials with low reflectivity are suitable for various anti-reflection coating in optical devices.

The refractive index (η) is an essential optical property for the solar cell absorber layer. A higher refractive index corresponds to higher reflectivity and lesser PCE for solar cells. Materials with a low value of η are very suitable for SCs absorber material as well as an anti-reflection layer. The refractive index of the thin films was calculated from the following equation [48],

$$\eta = \frac{1+R}{1-R} + \sqrt{\frac{4R}{(1-R)^2} - \left(\frac{\alpha\lambda}{4\pi}\right)^2}, \quad (6)$$

where R, λ, and α are the reflectance, incident wavelength, and absorption coefficient, respectively. **Figure- 11** shows the variation of η with increasing wavelength. The value of η is very low (1.48-1.67) for all the thin films. The refractive index of MAPI is comparatively low in the visible region and decreased after ZnS incorporation, reducing the energy loss due to reflection and improving the optical performance. For each thin film, the refractive index decreases with increasing wavelength in the visible region and then increases with increasing wavelength in the longer wavelength region. This suggests that longer wavelengths are reflected more from the materials in comparison to the visible wavelength. At higher wavelengths, η increases slightly with the increase of ZnS stoichiometry.

The bandgap of the films was calculated via Tauc relation [49] involving absorption coefficient (α) and photon energy given by the following equation,

$$\alpha h\nu = A(h\nu - E_g)^n \qquad (7)$$

where A, h, $\nu$, and $E_g$ are constant called band tailing parameter, Planck's constant, frequency of the photon, and bandgap, respectively. Tauc plot has plotted in **Figure- 12**, which revealed the bandgap MAPI thin film about 1.55 eV, satisfying the previous report [41]. The incorporation of ZnS caused to increase in the bandgap energy from 1.55eV to 1.64eV with increasing ZnS concentration. Obtained optical band gaps are shown in **Table 4**.

The XRD result shows that the lattice parameters and unit cell volume decreased with increasing the ZnS ratio in the MAPI samples. Due to the decrease of lattice parameters, molecules become more closely packed in the presence of ZnS; hence the electrons are more strongly bound in the lattice. Consequently, more energy will be required to free the electrons from the valance band (VB) due to the increase in the electron's binding energy with parent atoms, meaning the bandgap increases. The decrease of lattice parameter also represents the decrease in bond length. As a result, the energy gap between bonding orbital and antibonding orbital increases, which also indicates the increase of bandgap. It is also observed that the average crystallite size decreased from 83.02 nm to 53.5 nm, which can also be a reason for bandgap increment due to quantum confinement [50]. The obtained bandgaps are highly preferable for absorbing the visible light spectrum.

**Figure- 13** shows the optical conductivity (OC) variation with respect to incident photon energies for all prepared samples. The OC was calculated using the following equations [44],

$$\sigma = \frac{\alpha \eta c}{4\pi}, \qquad (8)$$

where σ, η, c, and α are optical conductivity, refractive index, speed of light, and thin-film absorption coefficient, respectively. Optical conductivity represents the rise of conductive property in the presence of light. Electron-hole pairs can be generated through the absorption of photon energy, which increases the electrons in CB and holes in VB. Hence the conductivity increases. In MAPI thin film, OC increased with the incident photon energy.

The incorporation of ZnS increased the absorption coefficient resulting in a boost of free carriers. The increase in free carriers can lead to the increase of OC of the thin films [51]. The variation of σ with photon energy showed an almost similar nature with a blue shifting of spectrum edge with the increase of ZnS stoichiometry. The OC blue-shifted due to the decrease of cell volume and crystallite size, which increase the bandgap. As a result, more photon energy is required for electron transition from VB to CB. All the films showed higher OC of the order of $10^{14}$ s$^{-1}$. The increased $\sigma$ suggest that the material, MAPI:ZnS is much potential material for various OE (e.g., SCs, photodetector, etc.) applications.

*3.6. Temperature Variation of Resistance*

**Figure- 14** shows the variation of the film's electrical resistance ($\mathcal{R}$) with respect to operating temperatures (T). Due to the negative temperature coefficient of resistivity (TCR), the resistance of the thin films decreased with increasing temperature, since with increasing temperature, more valance electrons gain energies to overcome the bandgap and become conduction electrons. This means with the increase in temperature, the carrier concentration also increases; hence the resistance decreases. The resistance increased with the ZnS molar concentration since the grain boundary resistance ($R_{gb}$) per unit length increased, reducing the cross transport of carriers [52].

The bandgap of a semiconductor can be measured from the temperature dependence of resistance. According to the band theory of solids, the relation between resistance and temperature is given by,

$$\mathcal{R}_T = \mathcal{R}_P e^{\frac{\varepsilon_g}{2kT}}, \qquad (9)$$

where, $\varepsilon_g$, k, T, $\mathcal{R}_T$ represent the energy gap between the top of VB and the bottom of CB, Boltzmann constant, Kelvin temperature, and electric resistance at temperature T. $\mathcal{R}_p$ is a temperature-independent resistance (constant). The above equation can be modified to

$$\varepsilon_g = 2k \frac{\Delta \ln(\mathcal{R})}{\Delta T^{-1}}. \qquad (10)$$

From the graph of ln($\mathcal{R}$) vs. T$^{-1}$ shown in **Figure-14,** the slope was determined and finally, from equation-10, the bandgaps were calculated and listed in **Table 4.** It can be observed that the calculated values of bandgaps via electrical means are quite analogous with the ones obtained by Tauc relation containing slight deviations. In both cases, the bandgap energies increased via ZnS incorporation.

**4. Conclusions**

This research has demonstrated the successful synthesis of MAPI and MAPI:ZnS MRs with different molar ratios via the spin coating method. Variation of ZnS molar ratio caused changes in morphological and optical properties of MAPI. The XRD and FTIR graphs show fine crystallinity and appropriate molecular vibrations present in the prepared samples. High absorption in the visible wavelength region with a tunable bandgap between 1.55eV-1.64 eV through varying ZnS stoichiometry in the MAPI was perceived. The bandgaps were obtained via the Tauc method as well as an electrical method which showed similar results. This tunability of bandgap can improve performance in various applications like solar cells, laser, LEDs, etc. All the observed properties, including very low reflectance between 1.48-1.64 and high absorption coefficient over $10^4$ cm$^{-1}$ in the visible and near-visible wavelength range proving that MAPI:ZnS MRs are potential candidates for different optoelectronics applications.

**Acknowledgment**

This research has been supported by the Condensed Matter Physics Lab, Wazed Miah Science Research Centre, and Bangladesh Council of Scientific and Industrial Research by providing the research accommodation and characterization techniques. We thank the National Science and Technology (NST) Fellowship, Bangladesh, for offering financial support for this research.

**Data availability**

The data used to support the findings of this study have not been made available because it is still being used in an ongoing research.

**Conflicts of interest**

The authors declare that there is no conflict of interest regarding the publication of this article.

# Tables

**Table 1: Lattice parameters of MAPI and MAPI:ZnS crystals**

| Material | Lattice parameters (Å) | | | Phase ($\alpha = \beta = \gamma = 90°$) | Cell volume (Å$^3$) |
|---|---|---|---|---|---|
| | a | b | c | | |
| MAPI | 6.3095 | 6.3095 | 6.280 | Tetragonal | 250.01 |
| MAPI: ZnS (1:0.025) | 6.300 | 6.300 | 6.2590 | Tetragonal | 248.42 |
| MAPI: ZnS (1:0.05) | 6.2941 | 6.2941 | 4.4720 | Tetragonal | 177.16 |
| MAPI: ZnS (1:0.1) | 4.4568 | 4.4568 | 6.2811 | Tetragonal | 124.76 |

**Table 2: Average crystallite size, lattice strain, dislocation density, and stacking fault of MAPI and MAPI:ZnS.**

| Material | Average crystallite size (nm) | Average lattice strain ($\times 10^{-3}$) | Dislocation density ($\times 10^{-4}$ nm$^{-2}$) | Stacking fault ($\times 10^{-3}$) |
|---|---|---|---|---|
| MAPI | 83.02 | 1.10 | 2.50 | 1.13 |
| MAPI: ZnS (1:0.025) | 73.80 | 1.18 | 2.54 | 1.15 |
| MAPI: ZnS (1:0.05) | 37.13 | 1.23 | 2.76 | 1.22 |
| MAPI: ZnS (1:0.1) | 52.50 | 1.36 | 4.17 | 1.44 |

**Table 3: Elemental atomic % and weight % measure via EDX spectroscopy**

| Elements | | Materials MAPI | MAPI:ZnS (1:0.025) | MAPI:ZnS (1:0.05) | MAPI:ZnS (1:0.1) |
|---|---|---|---|---|---|
| C | Weight % | 23.56 | 9.18 | 7.38 | 19.02 |
|   | Atomic % | 64.64 | 46.02 | 40.81 | 65.01 |
| N | Weight % | 8.74 | 4.08 | 3.51 | 4.05 |
|   | Atomic % | 20.56 | 17.53 | 16.64 | 11.87 |
| Pb | Weight % | 27.55 | 31.21 | 28.53 | 25.64 |
|    | Atomic % | 4.38 | 9.07 | 9.15 | 5.08 |
| I | Weight % | 40.15 | 54.24 | 58.58 | 48.5 |
|   | Atomic % | 10.43 | 25.73 | 30.66 | 15.69 |
| Zn | Weight % | --- | 0.81 | 1.33 | 1.87 |
|    | Atomic % | --- | 0.75 | 1.35 | 1.17 |
| S | Weight % | --- | 0.48 | 0.67 | 0.92 |
|   | Atomic % | --- | 0.9 | 1.39 | 1.18 |

**Table 4: Band gap variation in optical and electrical method.**

| Materials | Slope (β) | $\mathcal{E}_g$ (eV) | $E_g$ (eV) |
|---|---|---|---|
| MAPI | 8757.99 | 1.511 | 1.55 |
| MAPI:ZnS (1:0.025) | 8959.49 | 1.546 | 1.57 |
| MAPI:ZnS (1:0.05) | 9091.51 | 1.569 | 1.595 |
| MAPI:ZnS (1:0.1) | 9380.72 | 1.619 | 1.64 |

# Figures

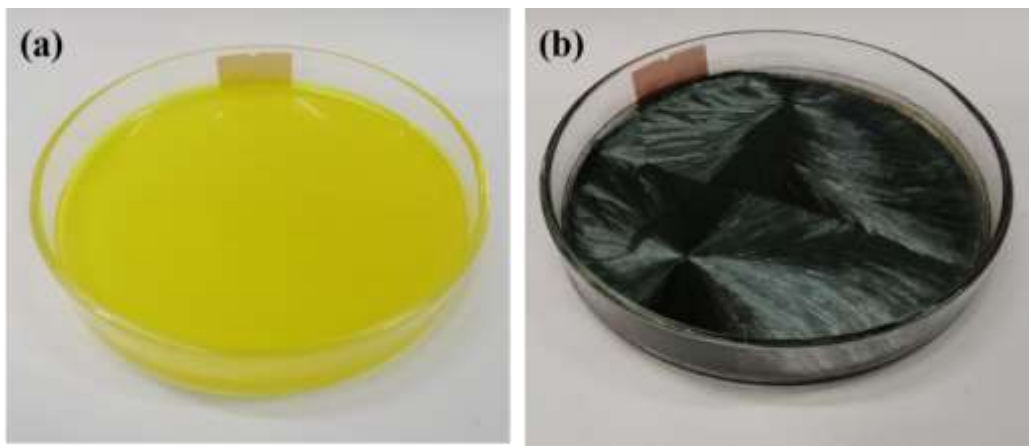

**Figure- 1: (a) MAPI solution in DMF, (b) MAPI dried crystals.**

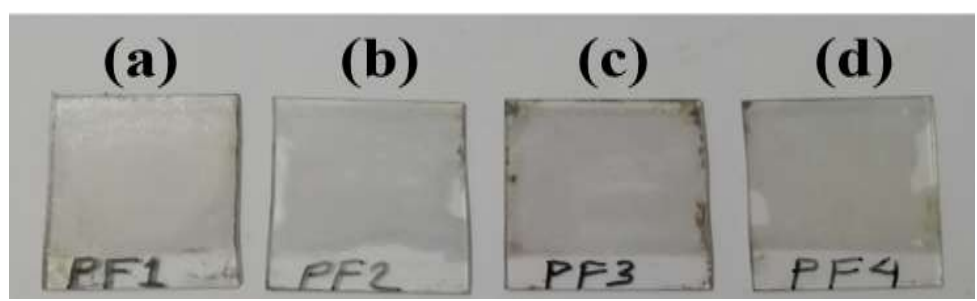

**Figure-2: Thin films of (a) MAPI, (b) MAPI:ZnS (1:0.025), (c) MAPI:ZnS (1:0.05), and (d) MAPI:ZnS (1:0.1) prepared by one-step spin coating method.**

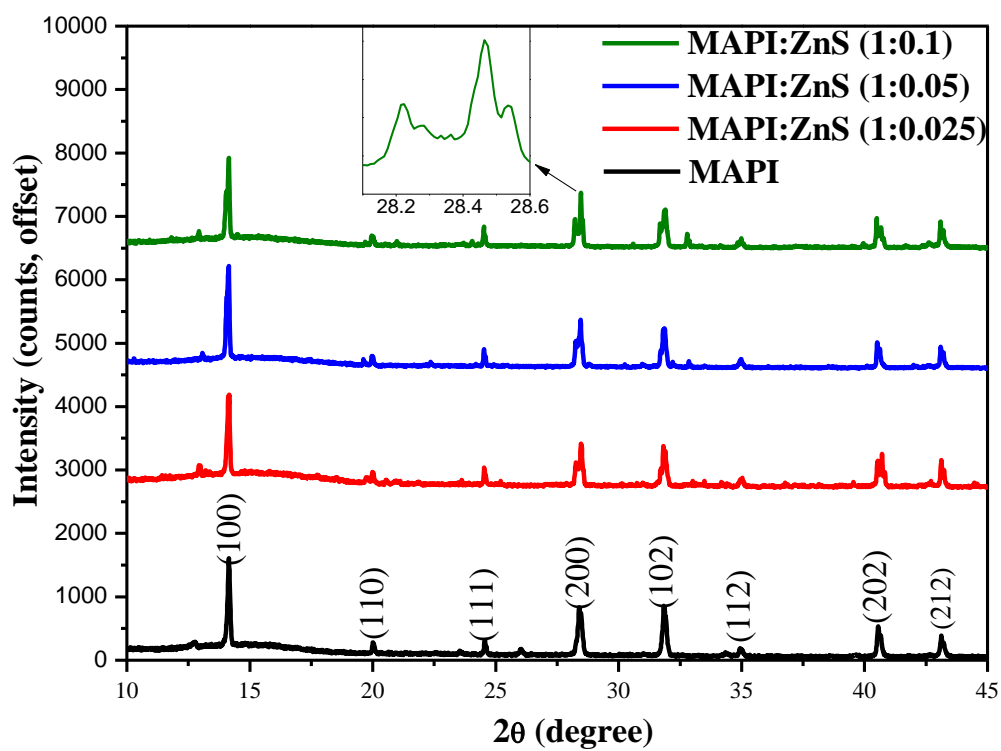

**Figure- 3: XRD spectra of MAPI and MAPI:ZnS thin films**

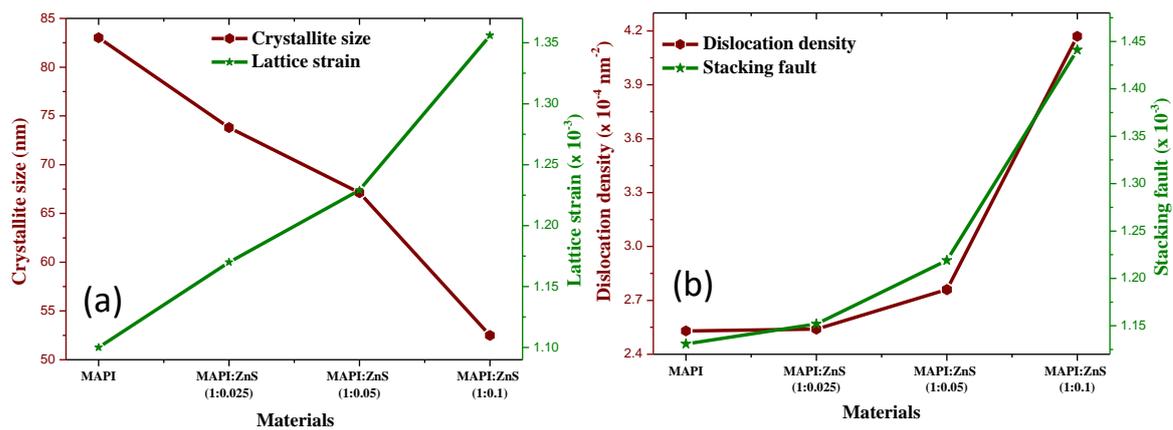

**Figure- 4: (a) Average crystallite size, lattice strain, and (b) dislocation density, stacking fault of MAPI and MAPI:ZnS.**

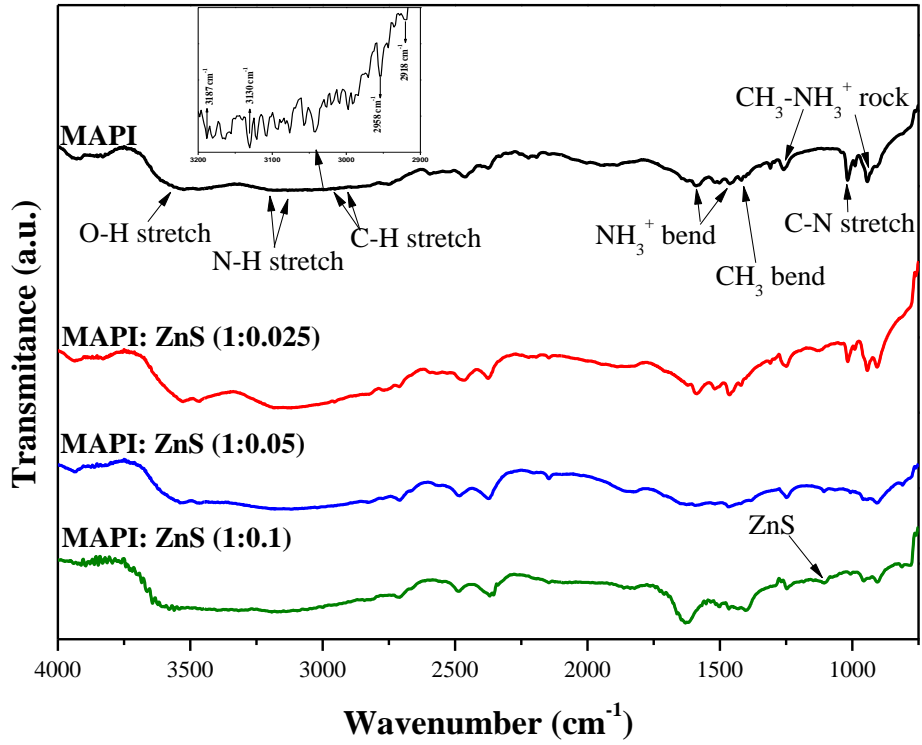

**Figure- 5: FTIR spectra of MAPI and MAPI: ZnS MRs**

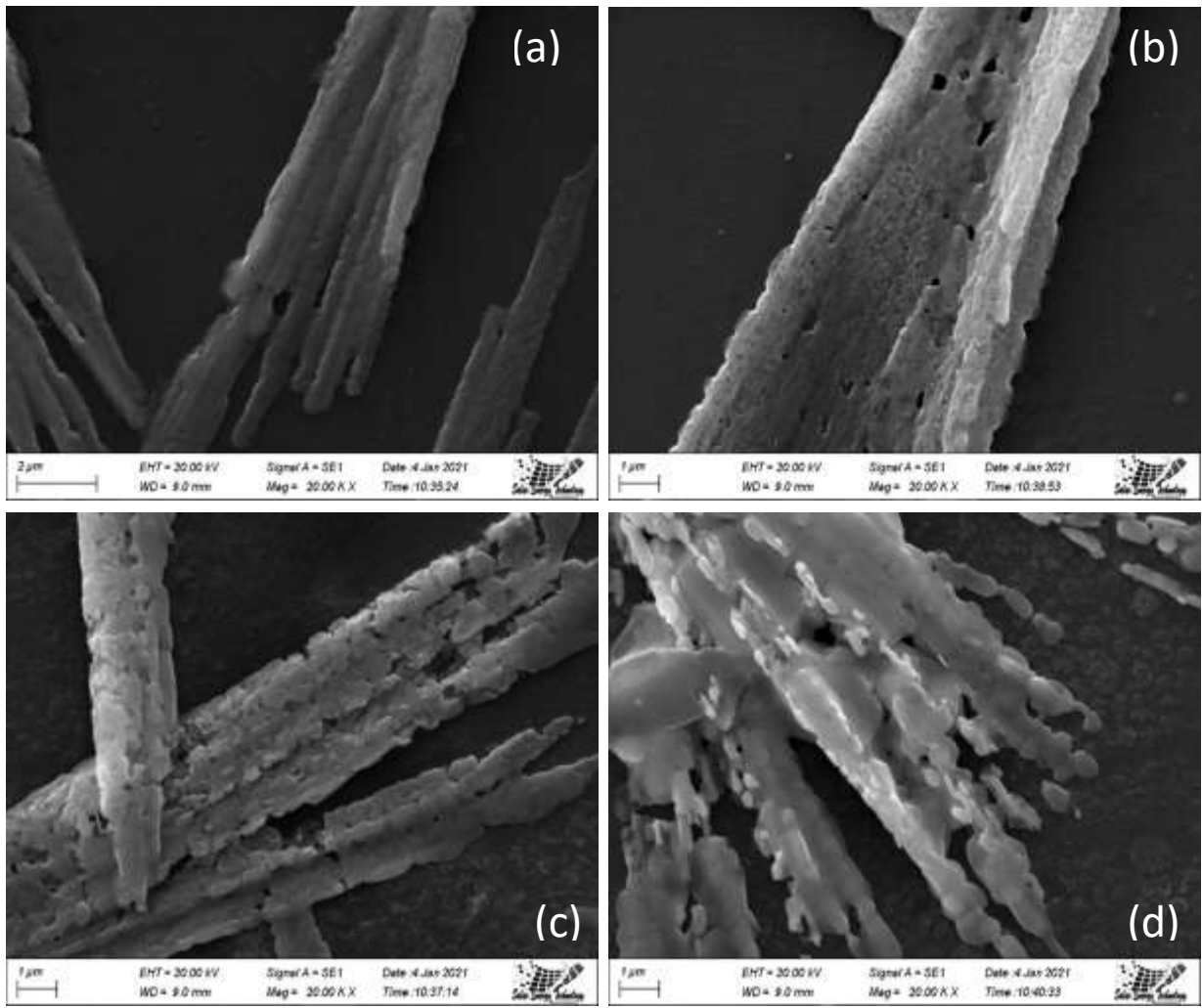

**Figure- 6: SEM images of (a) MAPI, (b) MAPI: ZnS (1:0.025), (c) MAPI: ZnS (1:0.05), and (d) MAPI: ZnS (1:0.1) thin films.**

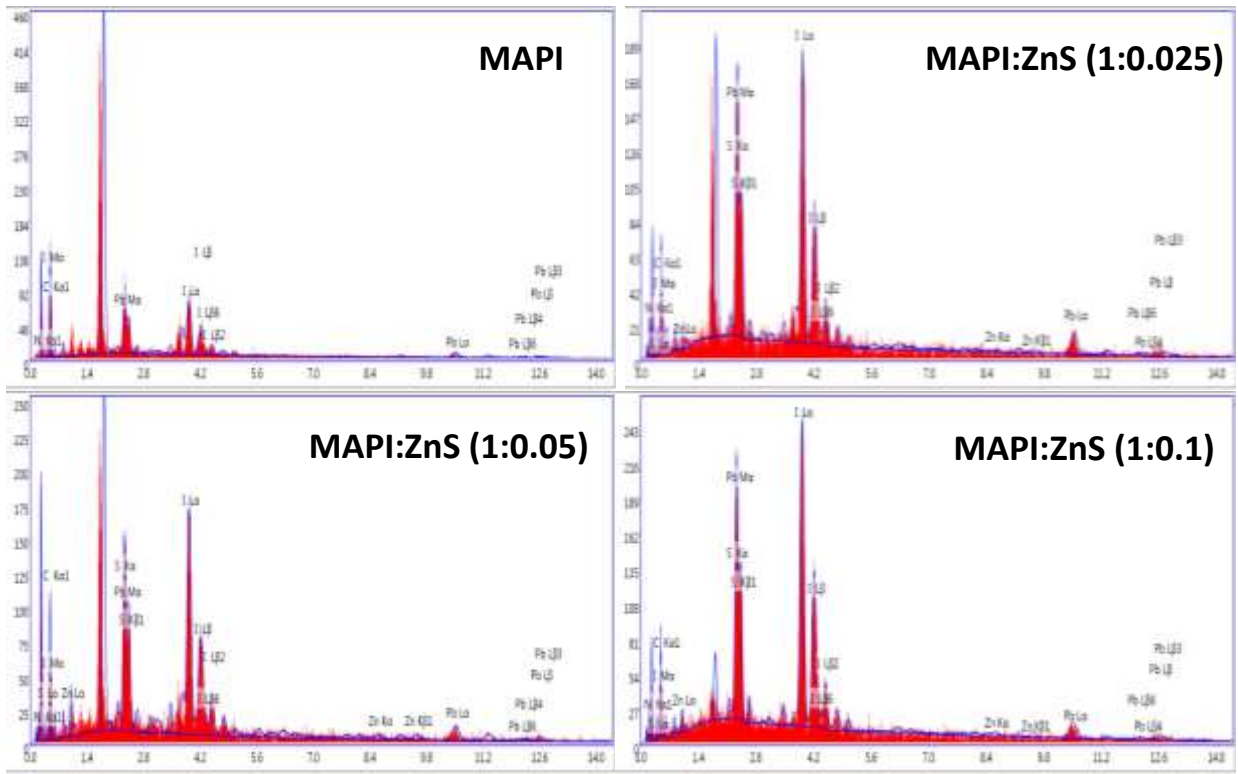

**Figure- 7: EDX spectra of the MAPI and MAPI:ZnS thin films.**

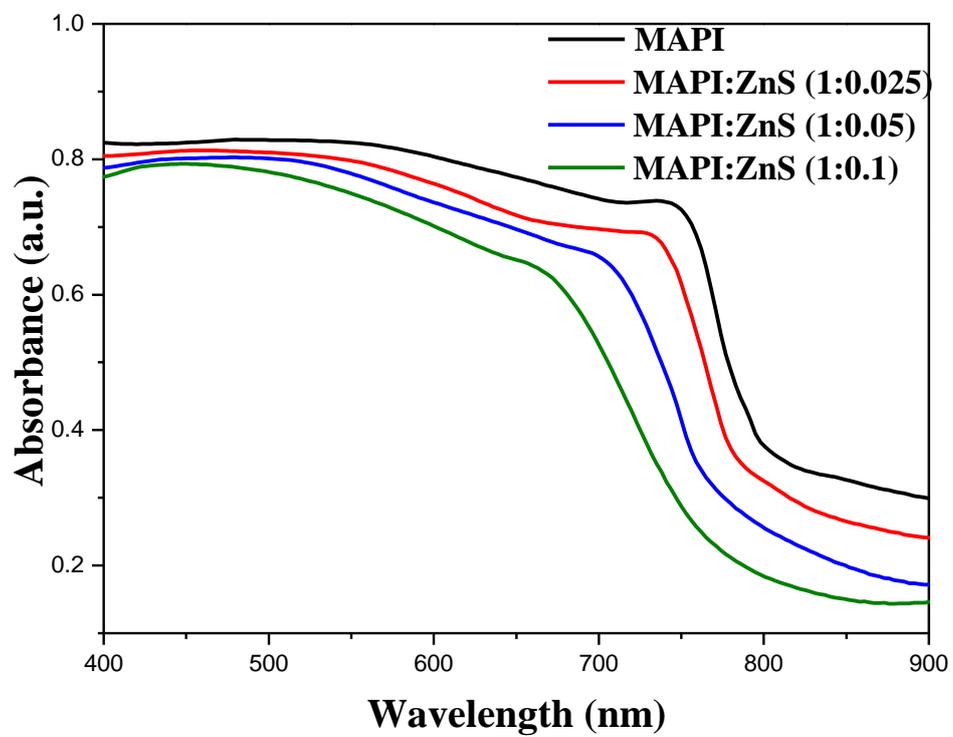

**Figure- 8: Absorbance spectra of the prepared thin films.**

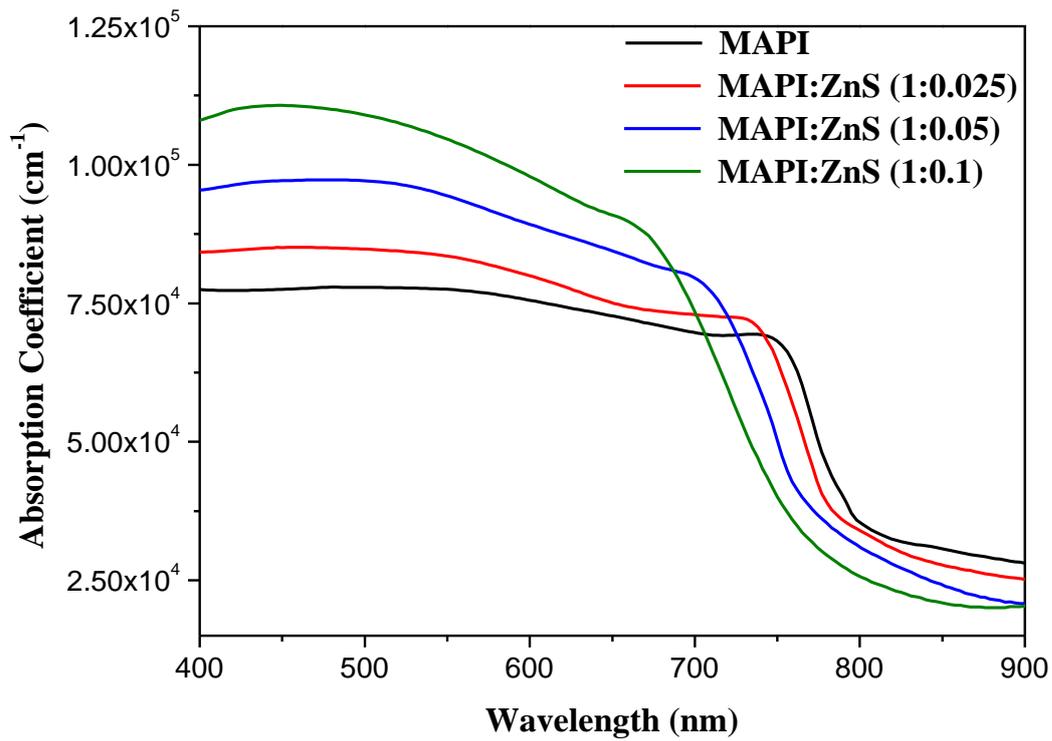

**Figure- 9: Absorption coefficients of the prepared thin films**

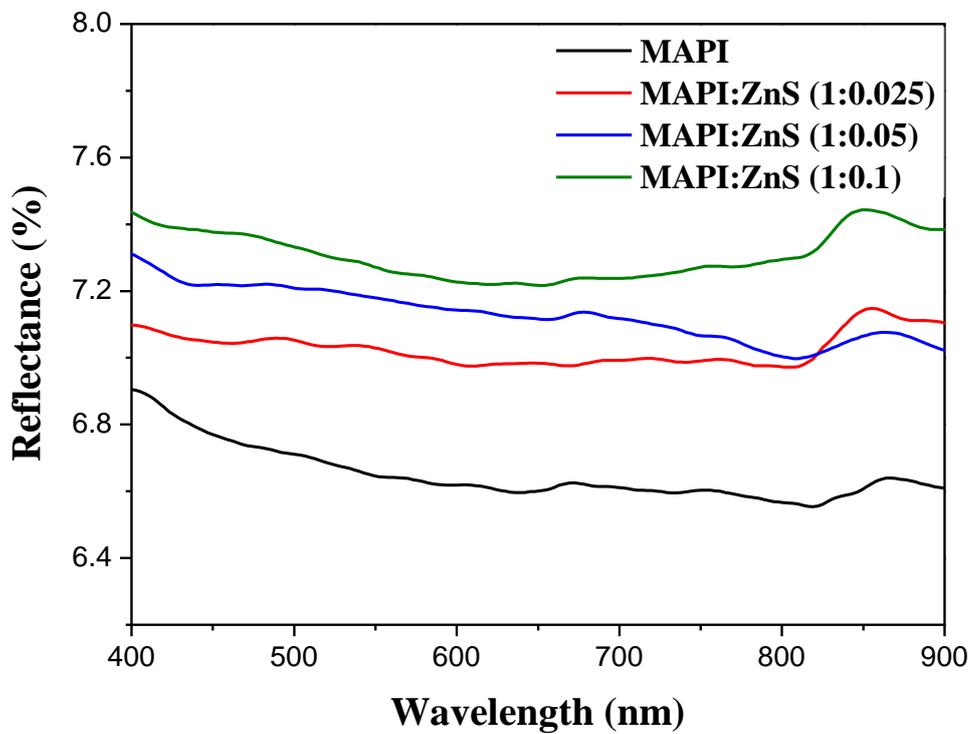

**Figure- 10: Reflectance spectra of the prepared thin films**

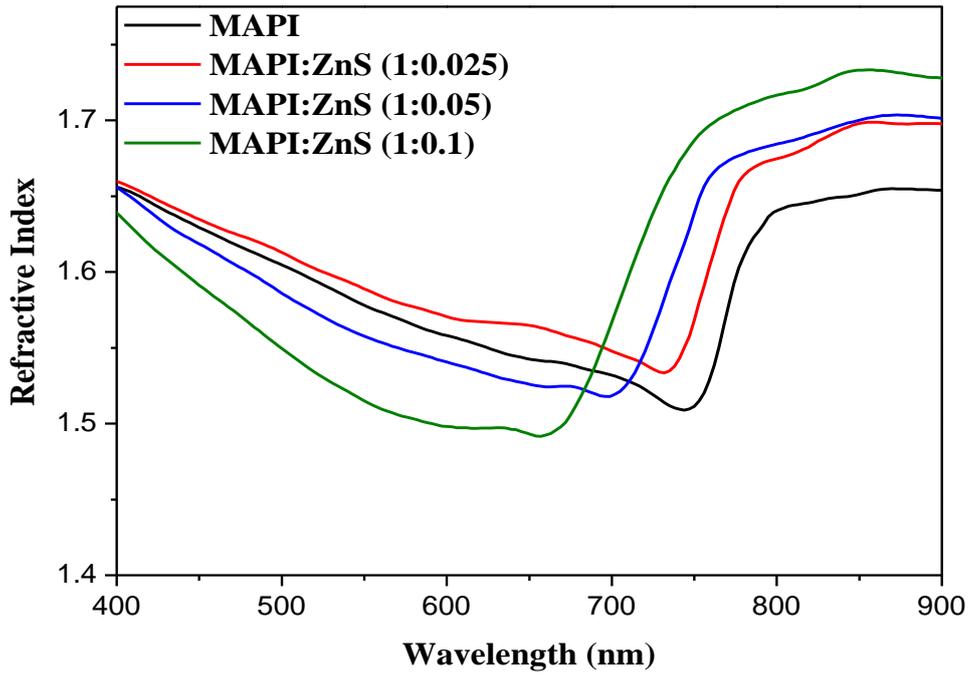

Figure- 11: Refractive index of the thin films

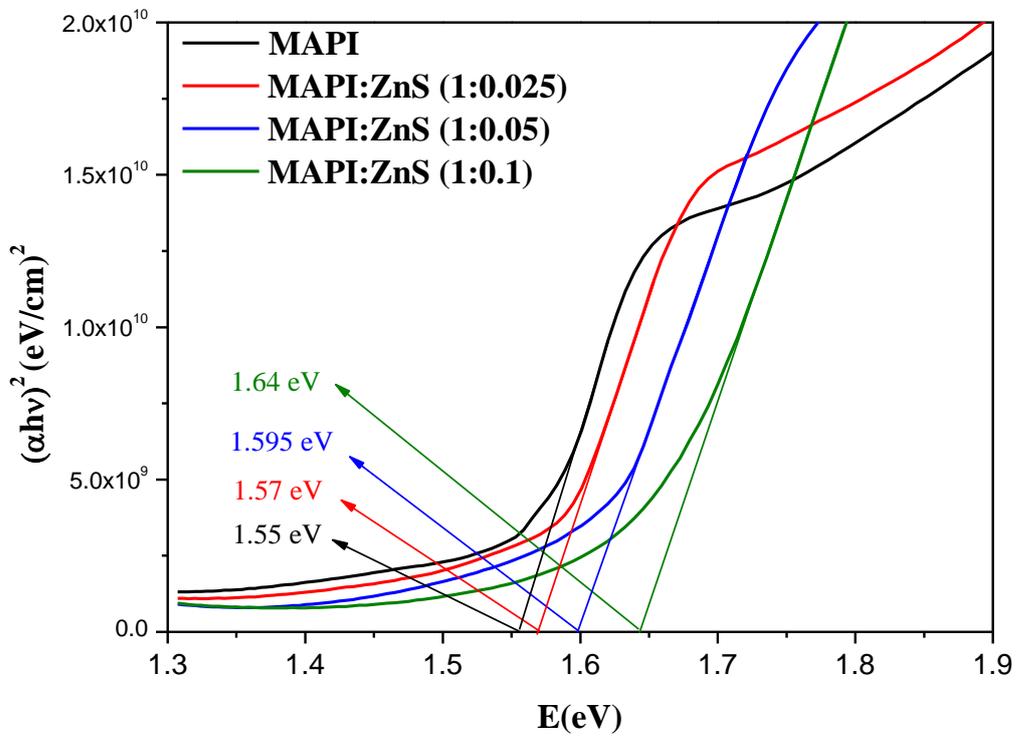

Figure- 12: Tauc plot for MAPI and MAPI: ZnS thin films

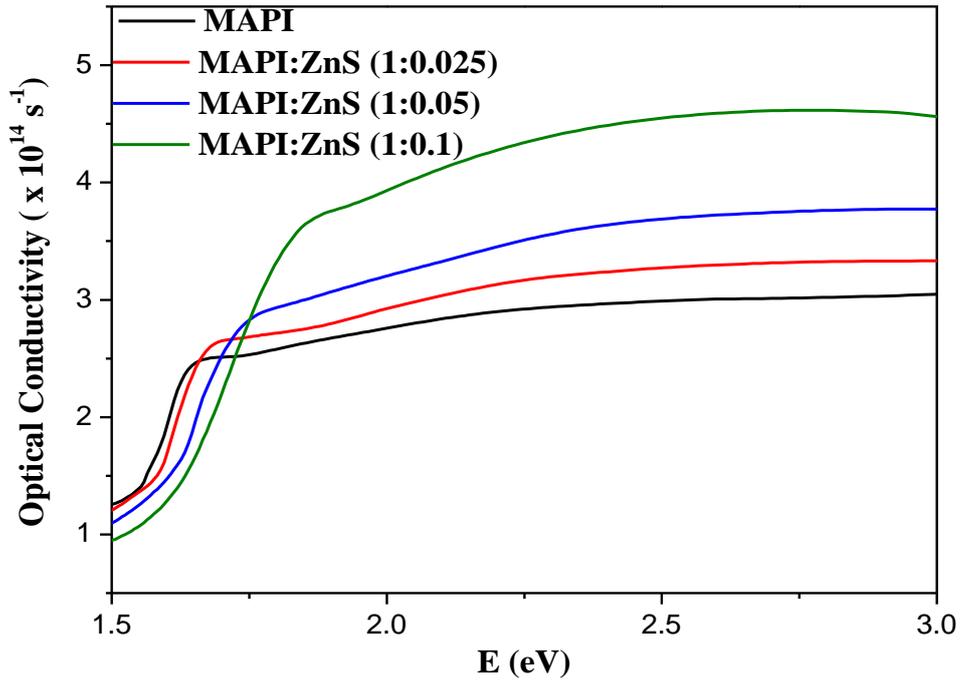

Figure- 13: Optical conductivity of the thin films

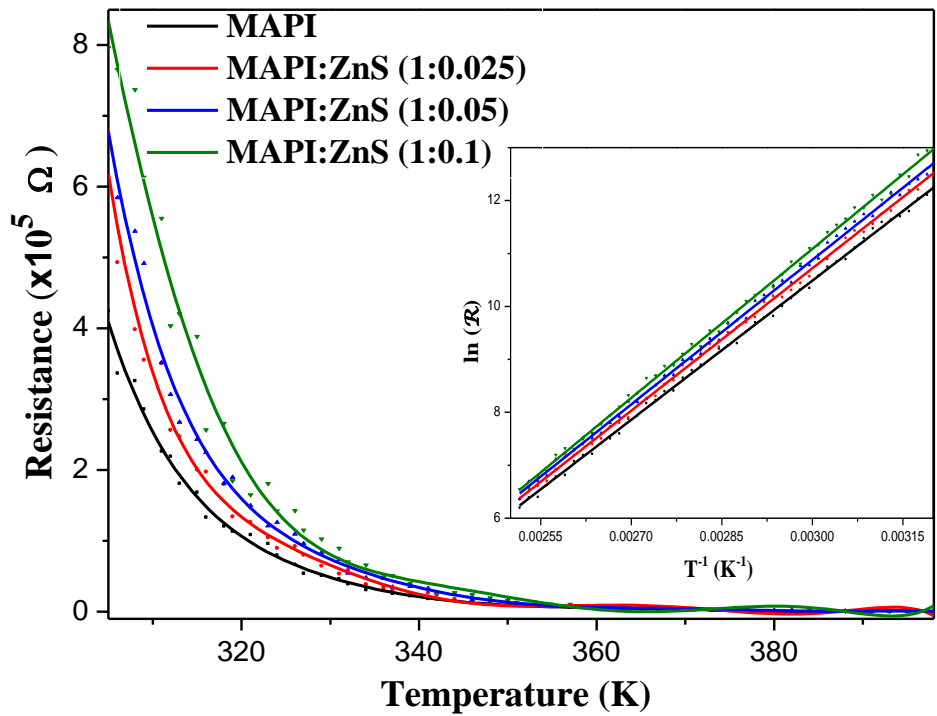

Figure- 14: Temperature variation of resistance of the prepared MAPI films.